\documentclass[]{JHEP3}
\title{Dirac Leptogenesis in extended nMSSM}

\author{
  Eung Jin Chun\\
 Korea Institute for Advanced Study, Hoegiro 87,
 Dongdaemun-gu, Seoul 130-722, Korea\\
 E-mail: \email{ejchun@kias.re.kr} }
\author{
 Probir  Roy\footnote{ DAE Raja Ramanna Fellow.}\\
 Saha Institute of Nuclear Physics, Block AF, Sector 1, Kolkata
 700 064, India\\
 E-mail: \email{probirr@gmail.com} }

\abstract{ We show that a version of the nearly Minimal
Supersymmetric Standard Model (nMSSM), extended only in the
singlet sector to include the additional superfields of
right-handed neutrinos and very heavy Dirac particles conserving
$B-L$,  admits a viable scenario for Dirac leptogenesis and
naturally small Dirac neutrino masses. The origin of the
($B-L$)-conserving high singlet neutrino scale and the desired
supersymmetry breaking terms is associated with dynamical
supersymmetry breaking in the hidden sector. }


\begin{document}

\section{Introduction}

Leptogenesis, generating a sphaleron-induced baryon asymmetry from
lepton asymmetry, can be realized not only with Majorana neutrinos
\cite{Fu86} but also with Dirac neutrinos conserving $B-L$
\cite{Di99}. Such neutrinos (as well as charged leptons) are first
produced in the early universe with equal numbers of left-handed
and right-handed states from the decays of very heavy fields.
However, the Yukawa couplings of the former are small enough to
hide, during the electroweak transition epoch, the right-handed
lepton number from sphalerons. This is as opposed to the
left-handed lepton number which gets converted to baryon number
through sphaleronic transitions.

\medskip

In the above context, a realistic supersymmetric model for Dirac
neutrino masses and leptogenesis was proposed in Ref.~\cite{Mu02},
and studied in detail in Refs.~\cite{Th05}. In this model, a
sub-eV neutrino mass scale gets generated by a combination of
reasonably small Yukawa couplings and a high mass scale of heavy
$SU(2)$ doublet superfields. Furthermore, leptogenesis can
successfully arise from the decays of such heavy fields. However,
there is a serious problem with this kind of a supersymmetric
scenario. The same heavy superfields can also decay into
right-handed sneutrinos which have masses of the order of hundreds
of GeV to a bit more or less than a TeV, as induced by soft
SUSY-breaking terms. Since their couplings $y L H N$, which are
supersymmetric generalizations of Yukawas, are tiny with $y \sim
10^{-13}$ guaranteeing the non-equilibration of the right-handed
neutrinos, they do not decay quickly enough and persist as relics
after the electroweak transition. In fact, their number density
divided by the entropy density then is expected to be as large as
$10^{-3}$. If they were stable, their relic density today would be
far too much in excess of that expected of Dark Matter (DM) as
estimated from cosmological observations. On the other hand,
because of their tiny couplings, unstable right-handed sneutrinos
can only decay very slowly, thus surviving till long after the
decoupling of the Lightest Supersymmetric Particle (LSP) which
takes place typically at a temperature of the order of 10 GeV.
Their late decays will then lead to a late non-thermal
over-production of the LSPs, once again conflicting with the
required DM relic density. Our aim is to present the conditions on
supersymmetry breaking parameters in the neutrino sector avoiding
this difficulty, and we present here a variant of the
Murayama-Pierce model \cite{Mu02} which achieves this goal. For
non-supersymmetric models for Dirac leptogenesis, see
Ref.~\cite{CE06}.

\medskip

Our spectrum extends from that of the Minimal Supersymmetric
Standard Model  \cite{Dr04} only in the sector that is a singlet
under the Standard Model gauge group. This extension is in the
spirit of the nearly Minimal Supersymmetric Standard Model nMSSM
\cite{Pa99}. In the latter, one adds to the MSSM superpotential
just one term, coupling a singlet chiral superfield $S$ to the two
Higgs superfields of the MSSM, in order to generate the
$\mu$-term.  In so far as direct effects on the low energy
superpotential are concerned, we need to add to the nMSSM spectrum
the right chiral neutrino superfields $N_i$, $i$ standing for
flavor,
%  a generic heavy singlet chiral superfield $X$ and
heavy Dirac pairs of singlet chiral superfields $\Phi_k, \Phi^c_k$
carrying $B-L$ charges $\pm1$,
%with $k$ being a type index.
and a generic heavy singlet superfield $X$ mediating supersymmetry
breaking. % from the hidden sector.
 In order to quicken the  decays of relic right-handed sneutrinos,
we need to have large values of certain supersymmetry breaking
parameters whose contributions to the scalar potential get scaled
by the neutrino mass and remain  small without much affecting the
lightest sparticle spectrum. Such large supersymmetry breaking
parameters may arise consistently from a dynamical supersymmetry
breakdown in the hidden sector. We show this explicitly  by
constructing a specific scheme in which the effects of such a
breaking are fed down to the observable (neutrino) sector by the
field $X$ associated with $U(1)_m$ mediation \cite{gmsb1,gmsb2}.

\medskip

In Section II we describe our mechanism for Dirac neutrino
leptogensis with a toy superpotential. In Section III we discuss
the required strengths of the supersymmtery breaking parameters to
avoid the late non-thermal overproduction of LSPs. The origin of
such parameters from a dynamically broken supersymmetry in the
hidden sector, fed down by $U(1)$ gauge mediation, is outlined for
a particular scheme in Section IV. Finally, Section V summarizes
our conclusions.

\section{Mechanism for Dirac neutrino leptogenesis}

Consider first the toy superpotential
 \begin{equation}
 W_{\rm Dirac}= h_{ik} L_i H_2 \Phi^c_k + h^\prime_{jk} N_j S \Phi_k  + M_k
 \Phi_k \Phi^c_k,
 \end{equation}
which will later constitute part of our full superpotential for
the observable sector. Here $h,h^\prime$ are Yukawa coupling
strengths and $M_k$ are heavy masses. We have used the standard
notation for the superfields $L_i, N_i, H_2$, etc. Integrating out
the heavy pairs, we have
 \begin{equation}
 W_{eff} =\, {h_{ik} h^\prime_{jk} \over M_k } \, L_i H_2 N_j S
 \equiv \, {m^\nu_{ij} \over v_2 v_S}\,  L_i H_2 N_j S,
 \end{equation}
generating the Dirac neutrino mass matrix  $m^\nu_{ij}$  with the
vacuum expectation values (VEVs) $v_2=\langle H_2^0 \rangle$ and
$v_S = \langle S \rangle$. Note that we need more than two pairs
of heavy fields ($\Phi_i ,\Phi^c_i $) in order to generate at
least two non-vanishing neutrino mass eigenvalues.   In this
approach, a tiny Dirac neutrino coupling $y = m^\nu / v_2 \sim
10^{-13}$ is obtained from reasonable values of the Yukawa
coupling strengths $h h^\prime \sim 10^{-6}$ and the heavy mass
scale $M_i \sim 10^{10}$ GeV with $v_S\sim$ TeV. At this point we
do not specify the origin of the VEV and the mass associated with
the singlet superfield $S$.

\medskip

Let us assume two pairs of heavy superfields  for simplicity,
which gives us two non-vanishing neutrino mass eigenvalues.  When
the masses of the heavy neutrinos are hierarchical, $M_1 \ll M_2$,
the decays of the lightest superfields $\Phi_1, \bar{\Phi}^c_1$
will produce the asymmetries $\epsilon$ and $-\epsilon$ in the
final states $L H_2$ and $ \bar{N}\bar{S}$, respectively. The
decay rate of $\Phi_1, \bar{\Phi}^c_1$ is
\begin{equation}
 \Gamma_D= {1\over 8\pi} \sum_i \left[ 2 |h_{i1}|^2 + |h^\prime_{i1}|^2 \right] M_1
\end{equation}
and the asymmetry $\epsilon$, defined by $\epsilon \,\Gamma_D
\equiv \Gamma(\Phi_1,\bar{\Phi}^c_1 \to L H_2) -
\Gamma(\bar{\Phi}_1,\Phi^c_1 \to \bar{L} \bar{H}_2)$, is
\begin{equation}
 \epsilon= { \sum_{i,j}\mbox{Im}(h_{i1}h^*_{i2} h^\prime_{j1} h^{\prime
 *}_{j2} ) \over 4\pi \sum_k ( 2 |h_{k1}|^2 + |h^\prime_{k1}|^2) }
 {\delta \over 1 - \delta^2},
\end{equation}
where $\delta\equiv M_1/M_2$ \cite{Th05}.  Here we assume that the
heavy masses $M_i$ are real and positive without loss of
generality. For a successful leptogenesis, the lepton asymmetry,
normalized by the entropy density $s$, $Y_L \equiv
(n_L-n_{\bar{L}})/s$, is required to be
\begin{equation}
 Y_L \approx \epsilon Y_\Phi \sim 10^{-10},
\end{equation}
where $Y_\Phi$ is the out-of-equilibrium density of the  heavy
superfield pair $\Phi_1, \Phi^c_1$. This is usually expressed in
terms of the ratio between the relativistic equilibrium number
density and the entropy density $n^{eq}_\Phi/s \sim 10^{-3}$ as
well as the efficiency factor $\eta \lesssim 1$: $Y_\Phi\equiv
(n^{eq}_\Phi/s) \eta$.  The efficiency factor, which is controlled
by the predominance of the decay over the inverse decay process,
is determined by the ratio  between the decay rate and the
expansion parameter of the universe $H(T)$ at $T=M_1$: $K\equiv
{\Gamma_D \over H(M_1) } \approx {0.1\over g_*^{1/2}} {M_P
 \over M_1} \sum_k (2 |h_{k1}|^2 + |h^\prime_{k1}|^2)$
where $M_P$ is the Planck mass and $g_*$ is the relativistic
degree of freedom at $T=M_1$. The approximate functional form of
the efficiency factor can be expressed as $\eta \approx 1/K (\ln
K)^{0.8}$ in the case of $K\gg 1$ \cite{Bu04}. From the relation
(2.2), one typically gets $K\sim (m_\nu /10^{-3} \mbox{eV})
(\mbox{TeV}/v_S)$ leading to $K\sim 10$ (and thus $\eta\sim 0.1$)
for the solar neutrino mass scale $m_\nu \sim 0.01$ eV and
$v_S\sim$ TeV. A more specific choice of parameters for this will
be shown below. To achieve the maximal efficiency $\eta\sim 1$
corresponding to $K\sim 1$, we may take $v_S \sim 10$ TeV or
introduce one more pair of heavy fields associated with a smaller
neutrino mass, $m_\nu \sim 10^{-3}$ eV, as in the case of the
usual seesaw mechanism \cite{Bu04}.

Let us remark that the asymmetry (2.4) is suppressed by the factor
$\delta\ll 1$ for hierarchical heavy neutrino masses.
Nevertheless, a realistic value of $\epsilon$ can be attained
\cite{Th05, Bu04} by choosing somewhat large magnitudes for
$h,h^\prime$. On the other hand, the CP asymmetry can be
resonantly enhanced if $\delta \approx 1$. This possibility  with
$h\sim h^\prime$ may well be motivated to generate the observed
mild hierarchy in the light neutrino masses. In this case, the CP
asymmetry is generalized to
\begin{equation}
 \epsilon= \sum_l { \sum_{i,j}\mbox{Im}(h_{i1}h^*_{i2} h^\prime_{j1} h^{\prime
 *}_{j2} ) \over 4\pi \sum_k ( 2 |h_{kl}|^2 + |h^\prime_{kl}|^2) }
 {\delta \over 1 - \delta^2},
\end{equation}
taking into account the contributions from the decays of two pairs
$\Phi_l, \Phi^c_l$ with $l=1,2$. The above formula  can be
straightforwardly generalized to the case of three families of
heavy superfields.   For a detailed analysis of the parameter
region satisfying the leptogenesis conditions, we refer the
readers to Refs.~\cite{Th05, Bu04}. As an example of the
parameters accommodating a successful leptogenesis, let us choose
$M_i \sim 10^{10}$ GeV and $h\sim h^\prime \sim 10^{-3}$.  This
gives us $K\sim 10$ leading to $\eta\sim 0.1$ and thus $Y_\Phi\sim
10^{-4}$ \cite{Bu04}. Now taking $1-\delta \sim 0.1$,  we get the
desired value of $\epsilon\sim 10^{-6}$.  Note also that one can
have almost degenerate heavy fields, $\delta\simeq 1$, in which
case the condition (2.5) can be met by significantly enhanced
$\epsilon$ and much smaller $Y_\Phi$.

Independently of the details of the parameter space, one can
conclude that successful leptogenesis via either (2.4) or (2.6)
implies $Y_\Phi>10^{-10}$.  This will lead in a supersymmetric
scenario to the problem of unwanted relics which we propose to
solve below.

%
%Successful leptogenesis (2.5) enforces the general condition
% \begin{equation}
%  Y_\Phi > 10^{-10}
%  \end{equation}
%in case of either (2.4) or (2.6).  This will lead to the
%problem of unwanted relics in the supersymmetric Dirac
%leptogenesis mechanism which we propose to solve below.
%

 \section{Conditions on
supersymmetry breaking parameters}

 A key requirement of Dirac leptogenesis is
not to equilibrate the right-handed neutrinos with left-handed
ones both above and at the electroweak phase transition when
sphaleron interactions are active. This is needed to make the
asymmetry in the left-handed neutrino turn into the baryon
asymmetry via those interactions \cite{Di99}. In other words, the
scattering rates $\Gamma_S$, induced by the effective operator
(2.2), needs to be suppressed as compared with the expansion rate
$H$ of the universe above and at the electroweak phase transition:
$\Gamma_S < H$ for $T \geq T_c$, where $T_c \sim 100$ GeV. For the
Dirac neutrino Yukawa operator $W= (m_\nu/v_2)\, L H_2 N$, this
condition requires $m_\nu \lesssim 10$ keV \cite{Di99} which is
trivially satisfied with the observed neutrino mass scale $0.05$
eV
 $\lesssim m_\nu \lesssim 0.33$ eV. Similarly, one can find that the
non-equilibration of the effective operator (2.2) itself is
satisfied for a temperature below $T_\nu$:
\begin{equation}
 T_\nu \sim 4\times 10^{15}
 \,\mbox{GeV} \left(   0.05\,
 \mbox{eV}\over m_\nu  \right)^2 \left(  v_S \over 1
 \,\mbox{TeV}\right)^{2}\,,
\end{equation}
which puts the bound; $M\lesssim T_\nu$, or $v_S \gtrsim 1.6$ GeV
for $m_\nu\gtrsim 0.05$ eV and $M\sim 10^{10}$ GeV.

The same argument has to be applied to the supersymmetry breaking
operators associated with the supersymmetric operators (2.2).
Taking
 general supersymmetry breaking parameters in the scalar
potential $V$, we write
\begin{equation}
 V= {m_\nu F_S \over v_2 v_S}\, \tilde{l} h_2 \tilde{n}
+ {\Lambda_\nu m_\nu \over v_2 v_S} \, \tilde{l} h_2 \tilde{n} s,,
\end{equation}
where $F_S$ is the $F$-term of $S$ and $\Lambda_\nu$ is a
supersymmetry breaking parameter which can come from the usual
$A$-term or $B$-term of the heavy field mass operator, as will be
shown below. For the moment, we do not specify the origins of
these supersymmetry breaking parameters keeping their values
arbitrary.   The requirement of not equilibrating the above
operators (3.2) leads to the conditions
\begin{equation}
 {F_S \over v_S} \lesssim 5\times10^7 \,\mbox{GeV}\quad\mbox{and}\quad
  {\Lambda_\nu \over v_S} \lesssim
 5\times 10^5  \,.
\end{equation}
With these bounds, we find the couplings in Eq.~(3.2) satisfying
$m_\nu F_S /v_2 v_S \lesssim 10^{-5}$ GeV and $m_\nu
\Lambda_\nu/v_2 v_S \lesssim 10^{-7}$. These values are too small
(compared with the weak scale) to lead to any observable
consequences on sparticle spectra and couplings probed in collider
experiments. In particular, it is impossible to realize the mixed
sneutrino dark matter \cite{Ar00,snuDM07} in the framework of
Dirac leptogenesis.

\medskip

%{\bf Problems with unwanted relics:}

Another important issue in supersymmetric Dirac leptogenesis is
the problem with unwanted relics.    During leptogenesis, the
out-of-equilibrium decays of the heavy fields ($\Phi_i,\Phi_i^c$)
produce the scalar components of $N$ and $S$ by the amount of
$Y_X$. Since their couplings are small as in Eqs.~(3.2, 3.3), they
never get equilibrated. Thus their initial abundances are retained
until they decay to the lightest supersymmetric particle (LSP)
which is considered to be dark matter.  If such a non-thermally
produced LSP survives, the observed dark matter abundance today
would put the upper bound
 \begin{equation}
 Y_X m_X \lesssim 10^{-10} {\rm GeV} \,.
 \end{equation}
This implies $Y_X \lesssim 10^{-12}$ for the LSP mass around 100
GeV which is clearly in contradiction to the condition (2.7) for
successful leptogenesis. To avoid this difficulty, we need to have
large couplings in Eq.~(3.2) to make the decay  of the scalar
$\tilde{n}$ occur while the LSP can equilibrate. Requiring the
decay temperature $T_D$ to be larger than the LSP decoupling
temperature $T_{LSP}$, which we take to be 10 GeV, we find
\begin{equation}
 \Lambda_\nu \;\;\mbox{or}\; \; {F_S\over v_S} \gtrsim 3\times10^6
 \left( \tilde{m}_N \over \mbox{TeV}\right)^{3 \over 4}
 \left( T_{LSP} \over 10 \,\mbox{GeV}\right)^{3 \over 4}  \,\mbox{GeV}.
\end{equation}
Similar bounds emerge from the scalar $s$ decay.  This
consideration is also applicable when the scalar $\tilde{n}$ (or
$s$) is the LSP. From Eqs.~(3.3, 3.5), the allowed range of the
supersymmetry breaking parameters is found to be $\Lambda_\nu$ or
$F_S/v_S \approx (10^6-10^8) \,\mbox{GeV}$. The challenge then is
to generate such a large supersymmetry breaking parameter. It
should be noted again that the corresponding contribution to the
scalar potential $V$ in Eq.~(3.2), being scaled by the tiny
neutrino mass, is controllably small and is not expected to give
rise to any undesirably large sparticle mass or any other significant
phenomenological consequence at laboratory energies. In the next
section, we will present a consistent framework for Dirac
leptogenesis realizing the conditions $F_S/v_S <$ TeV but
$\Lambda_\nu \sim 10^{7-8}$ GeV with $v_S\sim$ TeV.

\section{Origin of $v_S$ and $\Lambda_\nu$}

We now show that the conditions for successful Dirac leptogenesis
described above can be realized consistently in the context of a
suitably extended version of nMSSM. As in the usual version, let
us first couple the singlet superfield $S$ to the Higgs mass
operator through the superpotential term  $S H_1 H_2$ and thus
generate $v_S \sim \mbox{TeV}$ dynamically. In this scheme, there
is no problem with non-thermally produced $S$ as it can
equilibrate or decay fast by its large coupling with Higgs fields.
For a viable scenario of nMSSM containing the neutrino sector
(2.1), we introduce the following superfields and two global
symmetries $R$ and $B-L$ on top of the MSSM:
\begin{equation}
\begin{array}{|c|cccccccc|}
\hline
  & S & H_1 & H_2
  & L & N
  & \Phi & \Phi^c & X \\
  \hline
R & -{4\over3} & {5\over3} & {5\over3}
 & -{1\over3} & {8\over3}
 & {2\over3} & {2\over3} & {2\over3} \\
 \hline
B-L
  & 0 & 0 & 0
  & 1 & -1
  & -1 & 1 & 0 \\
  \hline
Y & 0 & -{1\over2} & {1\over2}
  & -{1\over2} & 0
  & 0 & 0 & 0 \\
  \hline
\end{array}
\end{equation}
Here we have taken the R-charge normalization such that $R(W)=2$ and $Y$ as the
Standard Model hypercharge. This allows us to write down the
superpotential
\begin{equation}
 W= \lambda S H_1 H_2 + h_{ik} L H_2 \Phi_k
 + h^\prime_{ik} N S \Phi^c_k
 + k_1 X \Phi \Phi^c + {k_2\over 3} X^3 + {k_3\over 5} {X^5\over M_P^3} S \,.
\end{equation}

It is now crucial for us to generate high scales for $v_X=\langle
X\rangle$ and $F_X$, which can be easily realized by adopting the
idea of dynamical supersymmetry breaking fed down by gauge
messengers \cite{gmsb1,gmsb2}.  As an illustration, let us
consider the $SU(6)\times U(1)\times U(1)_m$ model proposed in
Ref.~\cite{gmsb2}.  This model contains chiral superfields with
the quantum numbers as follows:
 $$
 {\cal A}\, (15,1,0),\;
 \bar{\cal F}^{\pm}\, (\bar{6}, -2, \pm1),\;
 {\cal S}^0\, (1, 3, 0),\;
 {\cal S}^{\pm2}\, (1, 3, \pm2).
 $$
 Without
contradicting our symmetry assignment (4.1), we can extend the
superpotential (4.2) to include the terms in Eqs.~(2.19), (2.23)
and (3.1) of Ref.~\cite{gmsb2}. Thus our dynamical supersymmetry
breaking (DSB) is described by the superpotential
\begin{equation}
 W_{DSB}= {\lambda_6\over M_P}  {\cal A} \bar{\cal F}^+ \bar{\cal F}^- {\cal S}^0
 + { \Lambda_6^5 \over
  ( {\cal A}\bar{\cal F}^+\bar{\cal F}^- {\cal A}^3)^{1/3} } +
 k_4 X \phi^+ \phi^- +
 {k_2\over3} X^3 \,,
\end{equation}
where $\lambda_6$ is a Yukawa coupling,  $\Lambda_6$ is the
dynamical scale of $SU(6)$, $\phi^{\pm}$ are the messenger fields
carrying the messenger hypercharge $\pm1$ under $U(1)_m$, and $X$
is the heavy singlet superfield used in Eq.~(4.2). The strong
dynamics of $SU(6)$ generates the second (non-perturbative) term
which drives dynamical supersymmetry breaking in combination of
the first (tree-level) term. Unlike the original propose for the
gauge mediated supersymmetry breaking \cite{gmsb2},  we will
assume a high scale $\Lambda_6 \sim 10^{13}$ GeV so that the
dynamically generated $F$-term $F_G\sim \lambda_6^{1/2}
\Lambda_6^{5/2}/M_P^{1/2}$ induces $\tilde{m} \sim F_G/M_P \sim$
TeV,  which is a right value for the gravitino mass and all the
soft masses of the MSSM through gravity mediation.  On the other
hand, dynamical supersymmetry breaking generates a negative
mass-squared for the $U(1)_m$-charged scalar field $\phi^\pm$, and
thereby $v_X$ and $F_X$ are induced through the last two terms of
Eq.~(4.3) as in Ref.~\cite{gmsb1,gmsb2}.

\medskip

Now we can show explicitly how the conditions (3.3, 3.5) are met
in our scheme. From the superpotential terms (4.2), one finds
\begin{equation}
 M=k_1 v_X ,\quad
 \Lambda_\nu= {F_X \over v_X}\quad\mbox{and}\quad
 F_S = {k_3\over 5} {v_X^5\over M_P^3} \,.
\end{equation}
One can easily adjust the Yukawa couplings in Eqs.~(4.2, 4.3) to
get, for instance,
\begin{equation}
 v_X \sim 10^{11}\, \mbox{GeV} \quad \mbox{and} \quad
 F_X/v_X \sim 10^8\,\mbox{GeV}\,,
\end{equation}
from which one obtains $\Lambda_\nu\sim 10^8$ GeV and $F_S \ll
\tilde{m}^2$.  The latter is too small to become relevant in
generating the vacuum expectation values of the Higgs bosons and
scalar $S$.  However, we can have in the scalar potential
\begin{equation}
 V = k_3 {v_X^4 F_X \over M_P^3} s \sim \tilde{m}^3 s
\end{equation}
which play the role of a tadpole contribution in the  nMSSM
\cite{Pa99}.  This term, together with the coupling $S H_1 H_2$,
gives rise to $v_S \sim$ TeV and thus  appropriate $\mu$ and
$B\mu$ terms respecting the Higgs stability and electroweak
symmetry breaking conditions \cite{Dr04}.  Let us finally note
that $\Lambda_\nu$ is in fact the $B$-term of the heavy mass
operator $M \Phi \Phi^c$. This makes it clear why we introduced
heavy singlets instead of $SU(2)$ doublets. If the heavy fields
$(\Phi, \Phi^c)$ were $SU(2)$ doublets as in Ref.~\cite{Mu02},
$\Lambda_\nu$ would induce via gauge mediation \cite{gmsb1,gmsb2}
too large soft supersymmetry breaking terms in the MSSM sector
characterized by $\tilde{m} \sim \alpha_2 \Lambda_\nu/4\pi$.

\section{Conclusion}

We have successfully implemented Dirac leptogenesis in an extended
version of the nMSSM. Unlike that of Murayama and Pierce
\cite{Mu02}, our extension of the MSSM involves only superfields
which are singlets under the Standard Model gauge group. The
advantage here is the quickening of the otherwise undesirably slow
decays of relic singlet sneutrinos, thereby avoiding a non-thermal
overproduction of LSPs which had plagued earlier Dirac
leptogenesis scenarios. This is achieved by generating heavy mass
scales $M\sim 10^{10}$ GeV and large supersymmetry breaking
parameters $\Lambda_\nu$ in the range $10^7 - 10^8$ GeV by
dynamical supersymmetry breaking.

Apart from the usual signatures of the Higgs sector extended with
singlet fields, our model has no signatures in low energy
phenomenology.  In particular,  the contribution of the seemingly
large parameter $\Lambda_\nu$ to the scalar potential get scaled
by the neutrino mass and is inconsequentially small in terms of
probing the scalar right-handed neutrino sector.  A clear
laboratory distinction between our model and that for standard
Majorana leptogenesis is that our light neutrinos are Dirac
particles with lepton number conserving interactions.  Thus,
non-observation of neutrinoless nuclear double beta decaying the
future experiments may hint at the Dirac nature of light
neutrinos, providing indirect support for Dirac leptogenesis.

\acknowledgments{ This investigation was initiated at the WHEPP X
workshop (Institute of Mathematical Sciences, Chennai, Jan.\ 2 --
13, 2008) and we thank the organizers for their hospitality. The
work of PR has been supported in part by the DAE BRNS of the
Government of India. }

\end{document}